\documentclass[pdflatex,sn-mathphys-num]{sn-jnl}

\usepackage{graphicx}%
\usepackage{multirow}%
\usepackage{amsmath,amssymb,amsfonts}%
\usepackage{amsthm}%
\usepackage{mathrsfs}%
\usepackage[title]{appendix}%
\usepackage{xcolor}%
\usepackage{textcomp}%
\usepackage{manyfoot}%
\usepackage{booktabs}%
\usepackage{algorithm}%
\usepackage{algorithmicx}%
\usepackage{algpseudocode}%
\usepackage{listings}%

\theoremstyle{thmstyleone}%
\theoremstyle{thmstyletwo}%

\theoremstyle{thmstylethree}%

\begin{document}

\title[Target-distribution-guided cross-functional MLIP fine-tuning]{Target-Distribution-Guided Cross-Functional Fine-Tuning of Machine-Learning Interatomic Potentials}


\author*[1,2]{\fnm{Yuki} \sur{Nagai}}\email{nagai.yuki@mail.u-tokyo.ac.jp}

\author[3]{\fnm{Bo} \sur{Thomsen}}

\author[3]{\fnm{Motoyuki} \sur{Shiga}}

\affil*[1]{
\orgdiv{Information Technology Center}, \orgname{The University of Tokyo}, \orgaddress{\street{6-2-3 Kashiwanoha},
\city{Kashiwa}, \state{Chiba} \postcode{277-0882}, \country{Japan}}}

\affil[2]{
\orgdiv{Department of Advanced Materials Science}, \orgname{The University of Tokyo}, \orgaddress{\street{5-1-5 Kashiwanoha}, \city{Kashiwa}, \state{Chiba} \postcode{277-8561}, \country{Japan}}}

\affil[3]{
\orgdiv{CCSE}, \orgname{Japan Atomic Energy Agency}, \orgaddress{\street{178-4-4, Wakashiba}, \city{Kashiwa}, \state{Chiba},
\postcode{277-0871}, \country{Japan}}}


\abstract{
Cross-functional fine-tuning of machine-learning interatomic potentials (MLIPs) is often treated as a relabeling problem, where configurations generated at one density-functional level are relabeled using a higher-fidelity target functional. However, the resulting training data may be drawn from the wrong equilibrium distribution, because the statistical weights of configurations change across exchange--correlation functionals. Here we address this distribution mismatch using a target-distribution-guided workflow based on self-learning hybrid Monte Carlo (SLHMC), in which trial configurations are proposed by a machine-learning potential and accepted or rejected using target-functional density-functional-theory energies. Using rutile TiO$_2$ as a test system, we fine-tune the MACE-MP-0 foundation potential toward PBE, r$^2$SCAN, and HSE06 target functionals. The resulting adapted potentials reproduce target-anchored nearest-neighbor Ti--O distributions, radial distribution functions, and the NPT cell metrics examined here more accurately than the foundation-model and off-target relabeling controls considered in this work. In particular, HSE06-guided fine-tuning improves structural and thermodynamic properties that are difficult to access with direct hybrid-functional molecular dynamics because of the computational cost of exact exchange. These results indicate that target-distribution coverage is an essential component of cross-functional MLIP transfer, and that accurate target-level labels alone may be insufficient when the configurational distribution is mismatched.
}

\keywords{machine-learning interatomic potentials, self-learning hybrid Monte Carlo, density functional theory, MACE, TiO$_2$, sampling distribution}



\maketitle

\section{Introduction}\label{sec1}

Machine-learning interatomic potentials (MLIPs) enable molecular simulations with near density functional theory (DFT) accuracy at much lower cost. Early high-dimensional neural-network potentials established the basic MLIP framework for representing DFT potential-energy surfaces \cite{Behler2007-hdnnp}. Recent equivariant models such as NequIP and MACE have improved accuracy and data efficiency, while foundation models trained on broad materials datasets can be adapted to specific systems through fine-tuning \cite{Batzner2022-ns,NEURIPS2022_4a36c3c5,Batatia2025-vv,Radova2025-iw}. This strategy is especially attractive when the desired target level of theory is too expensive for routine molecular dynamics.

Fine-tuning MLIPs across exchange--correlation functionals remains difficult. Within Jacob's ladder of DFT \cite{Perdew2001-jacob}, generalized-gradient approximations, meta-GGAs, and hybrid functionals form a hierarchy of increasing physical complexity and computational cost. Moving up this ladder changes not only the energy reference but also the potential energy surface, equilibrium structures, and statistical weights of configurations. Energy-alignment strategies can stabilize cross-functional transfer \cite{Huang2025-zw}, but a model with low energy and force errors can still sample an incorrect equilibrium ensemble.

Here we focus on the sampling distribution of the training configurations. Conventional cross-functional workflows generate configurations using lower-level DFT or lower-level MLIPs and then relabel them at the target level \cite{Radova2025-iw,Huang2025-zw}. Recent cross-functional transfer studies have likewise reported that transfer toward higher-fidelity targets such as r$^2$SCAN can remain difficult even when target-level labels are available, indicating that label fidelity alone does not remove all transfer barriers \cite{Huang2025-zw}. The training set is therefore effectively drawn from a lower-level distribution, whereas the target thermodynamic observables are defined by the Boltzmann distribution of the target functional,
\begin{align}
    P_{\mathrm{target}}(\mathbf{R}) \propto \exp[-\beta E_{\mathrm{target}}(\mathbf{R})],
\end{align}
where $\mathbf{R}$ denotes the atomic configuration. This distribution mismatch is especially severe for hybrid functionals such as HSE06, for which direct ab initio molecular dynamics is costly because of exact exchange \cite{Heyd2003-sy,Krukau2006-ii}. Configurations that are important under the target functional may be underrepresented or absent, and cannot be expected to be reliably recovered simply by relabeling, random subsampling, or selecting configurations from the same biased trajectory.

This is the atomistic analogue of distribution shift in machine learning: a model fine-tuned on data drawn from one distribution can perform poorly when deployed on another, even when the labels themselves are accurate \cite{Kumar2022-sj,Wortsman2021-nf}. In MLIPs, the deployment distribution is the equilibrium ensemble generated by the learned potential, so the sampling distribution used for fine-tuning directly affects thermodynamic observables.

Rutile TiO$_2$ provides a particularly informative test case for this problem. It is a widely used benchmark system for oxide MLIPs, its Ti--O coordination environment and equilibrium cell parameters are sensitive to the choice of exchange--correlation functional, and hybrid-functional sampling is substantially more expensive than semilocal sampling because of exact exchange. The system is also representative of chemically relevant transition-metal oxides used in photocatalysis, oxide interfaces, and defect studies. It is therefore a meaningful case in which to test whether target-distribution-guided transfer improves the structural ensemble sampled by a fine-tuned potential.

We address this mismatch using self-learning hybrid Monte Carlo (SLHMC) \cite{Nagai2020-gx,Thomsen2024-dn}. Trial configurations are proposed by the current MACE potential, but accepted or rejected using target-functional DFT energies. For a fixed proposal potential, this construction anchors the Markov chain to the target Boltzmann distribution. The DFT-labeled proposal stream, including accepted and rejected configurations, is then used to update the active MACE model during sampling. For rutile TiO$_2$, we fine-tune MACE-MP-0 toward PBE, r$^2$SCAN, and HSE06. We show that SLHMC-based fine-tuning reproduces nearest-neighbor distributions, radial distribution functions, and equilibrium cell properties more accurately than lower-level and foundation-model sampling controls, and that sampling distribution is an important factor in cross-functional MLIP transfer.

\section{Results}\label{sec:results}

\subsection{SLHMC anchors fine-tuning to target-distribution sampling}

We begin by clarifying the distinction between the distribution used for error minimization and the equilibrium ensemble that the final potential is expected to reproduce. Fine-tuning minimizes a supervised loss over the distribution from which training configurations are sampled,
\begin{align}
    \theta^{*}
    =
    \arg\min_{\theta}
    \mathbb{E}_{\mathbf{R}\sim P_{\mathrm{train}}}
    \left[
    \mathcal{L}
    \left(
    E_{\theta}(\mathbf{R}), \mathbf{F}_{\theta}(\mathbf{R});
    E_{\mathrm{target}}(\mathbf{R}), \mathbf{F}_{\mathrm{target}}(\mathbf{R})
    \right)
    \right],
\end{align}
where $\theta$ denotes the parameters of the MLIP. In conventional relabeling workflows, the labels are evaluated at the target level of theory, but the sampling distribution remains that of the lower-level or foundation-model dynamics, $P_{\mathrm{train}} \simeq P_{\mathrm{base}}$. Thus the optimized potential is biased toward reducing errors on configurations typical of the base model, rather than on configurations typical of the target ensemble, $P_{\mathrm{target}}$.

In SLHMC, $P_{\mathrm{train}}$ is no longer set by an unrelated lower-level trajectory. Trial configurations are proposed using the current machine-learned potential, but the Markov chain is accepted or rejected using target-functional DFT energies. For each fixed proposal potential, the HMC kernel targets the Boltzmann distribution of the DFT functional used in the acceptance test. Because the proposal potential is updated during self-learning, the accumulated dataset is best viewed as target-distribution-guided data generation rather than as a single equilibrium trajectory from a stationary kernel.

This limitation is not removed by replacing random subsampling with more elaborate data selection. Such procedures can change which points are selected from a trajectory, but they cannot recover configurations that the lower-level dynamics does not visit.

\begin{figure}[htbp]
\centering
\includegraphics[width=\textwidth]{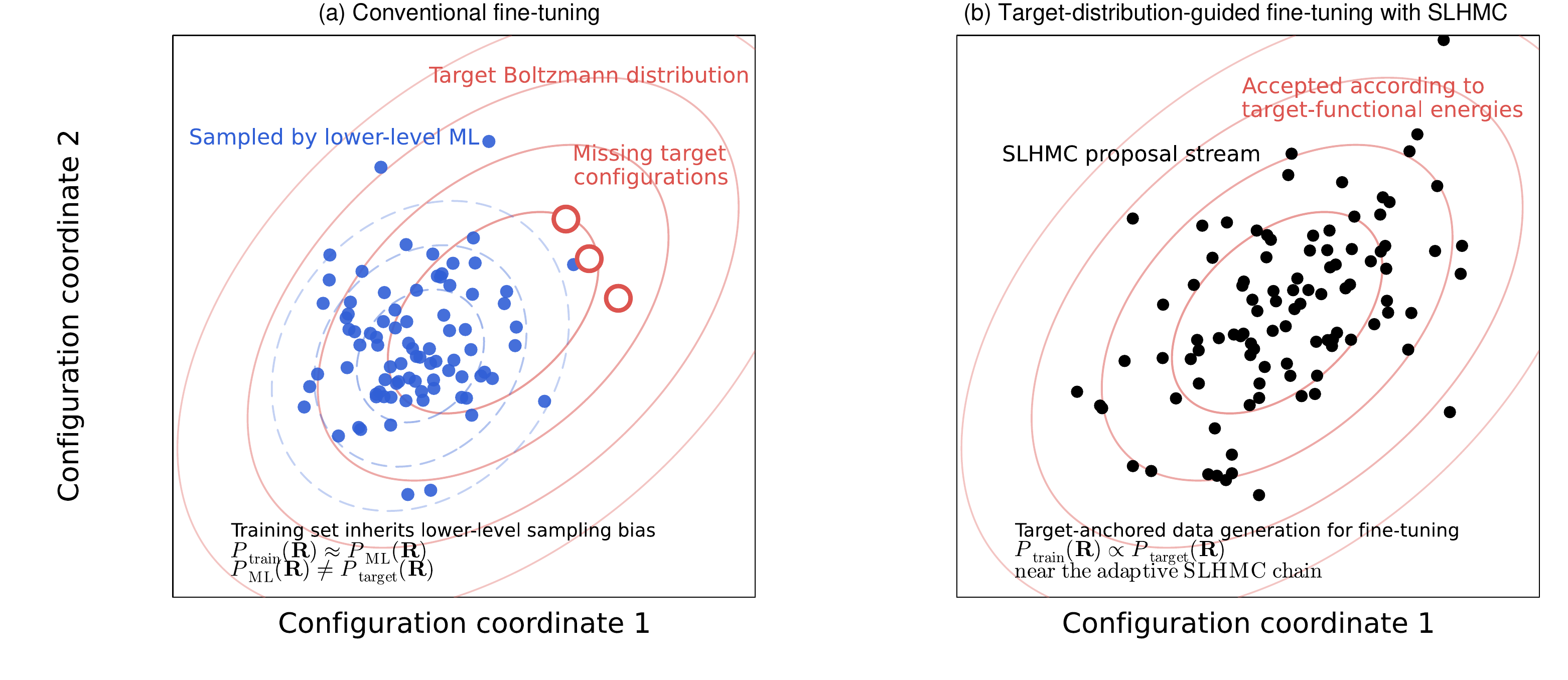}
\caption{Schematic view of distribution mismatch in conventional fine-tuning and its mitigation via SLHMC. (a) In conventional fine-tuning workflows, atomic configurations are generated using lower-level potentials, such as GGA-based MD, so the sampling distribution can differ from the target-functional Boltzmann distribution. Important configurations may then be underrepresented or missing. (b) In SLHMC, trial configurations are accepted or rejected using target-level DFT energies, and the resulting proposal stream is used for target-distribution-guided fine-tuning.}
\label{fig:concept}
\end{figure}

In the present implementation, the supervised dataset accumulated during SLHMC consists of target-functional evaluations along the proposal stream and therefore includes both accepted and rejected configurations. It is thus distinct from the accepted-state ensemble used later as a structural reference, but it remains tied to target-functional acceptance rather than to an off-target lower-level trajectory. Figure~\ref{fig:concept} summarizes this distinction.

\subsection{Forward prediction errors improve during SLHMC fine-tuning}

\begin{figure}[htbp]
\centering
\includegraphics[width=0.74\textwidth]{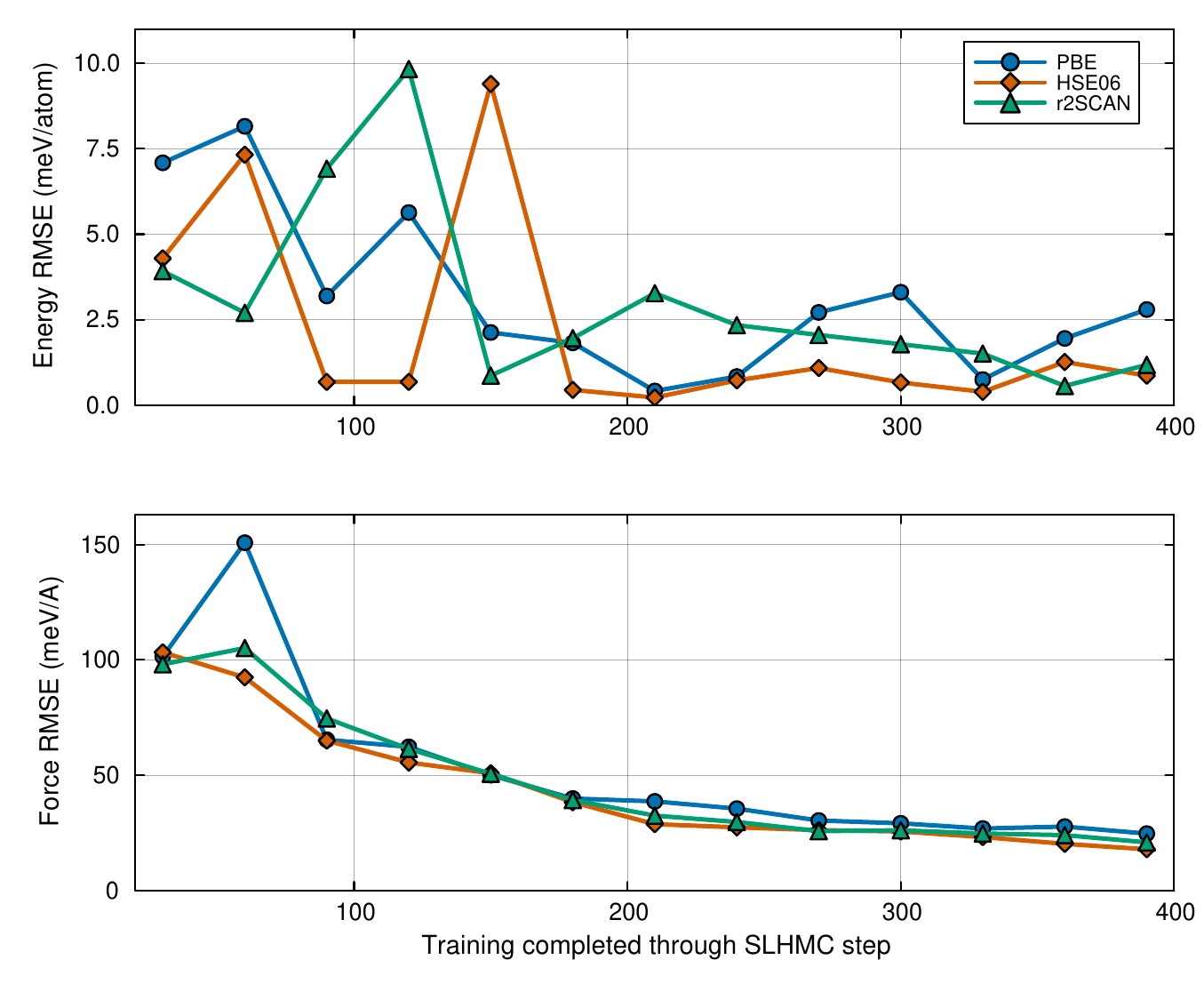}
\caption{Evolution of forward prediction errors during SLHMC-based fine-tuning. Energy and force RMSE values are shown as a function of completed SLHMC steps for the PBE, r$^2$SCAN, and HSE06 target functionals. All three series are plotted through 390 steps for a common comparison window.}
\label{fig:forward-rmse}
\end{figure}

We next examine whether the SLHMC self-learning loop yields a practically useful surrogate potential as sampling proceeds. Figure~\ref{fig:forward-rmse} reports forward-in-time energy and force RMSE values, where a model trained using data accumulated up to step $t$ is evaluated on subsequent target-functional configurations.

For the three target functionals included in this forward-error analysis, PBE, r$^2$SCAN, and HSE06, force errors decrease substantially as additional SLHMC-generated data are incorporated. The initial force RMSE values are about 98--103 meV \AA$^{-1}$ after the first 30-cycle update and decrease to 32 meV \AA$^{-1}$ for PBE, 16 meV \AA$^{-1}$ for r$^2$SCAN, and 15 meV \AA$^{-1}$ for HSE06 at the final analyzed updates. Energy errors are smaller in absolute magnitude but less monotonic, with occasional spikes such as the HSE06 value at 150 cycles. We therefore interpret Fig.~\ref{fig:forward-rmse} as evidence of practical model improvement rather than as a strictly monotonic convergence curve or as an independent benchmark of global model accuracy.

The iterative update is stable across these levels of electronic structure theory, and the foundation MACE model can be adapted using data generated within the SLHMC loop. At the same time, RMSE reduction alone does not guarantee that the learned potential reproduces the correct equilibrium ensemble. A model may achieve lower average prediction error while still misrepresenting the statistical weight of thermodynamically important configurations. We therefore next evaluate structural observables that are directly sensitive to sampling distributions.

The sampling process itself also remains stable throughout the analyzed runs. The NVT SLHMC simulations reach cumulative acceptance ratios of about 0.69--0.75 by the end of the analyzed trajectories, whereas the NPT runs settle to lower but still steady values of 0.26 for PBE and 0.19 for HSE06 because cell fluctuations make the proposals more demanding. The corresponding acceptance histories and adapted proposal lengths are summarized in Appendix Fig.~\ref{fig:hmc-adaptation}.

\subsection{Nearest-neighbor distributions expose sampling mismatch}

The impact of the sampling protocol becomes apparent when examining the nearest-neighbor Ti--O distance distribution. This quantity provides a sensitive probe of local structure, as it reflects short-range coordination environments that vary significantly across different exchange--correlation functionals. The trajectory labels used in the structural comparisons are defined in Appendix Table~\ref{tab:protocols}; in brief, ``MACE MD'' denotes trajectories generated with a frozen potential, whereas ``SLHMC accepted states'' denotes configurations accepted by target-functional DFT energies.

\begin{figure}[htbp]
\centering
\includegraphics[width=0.86\textwidth]{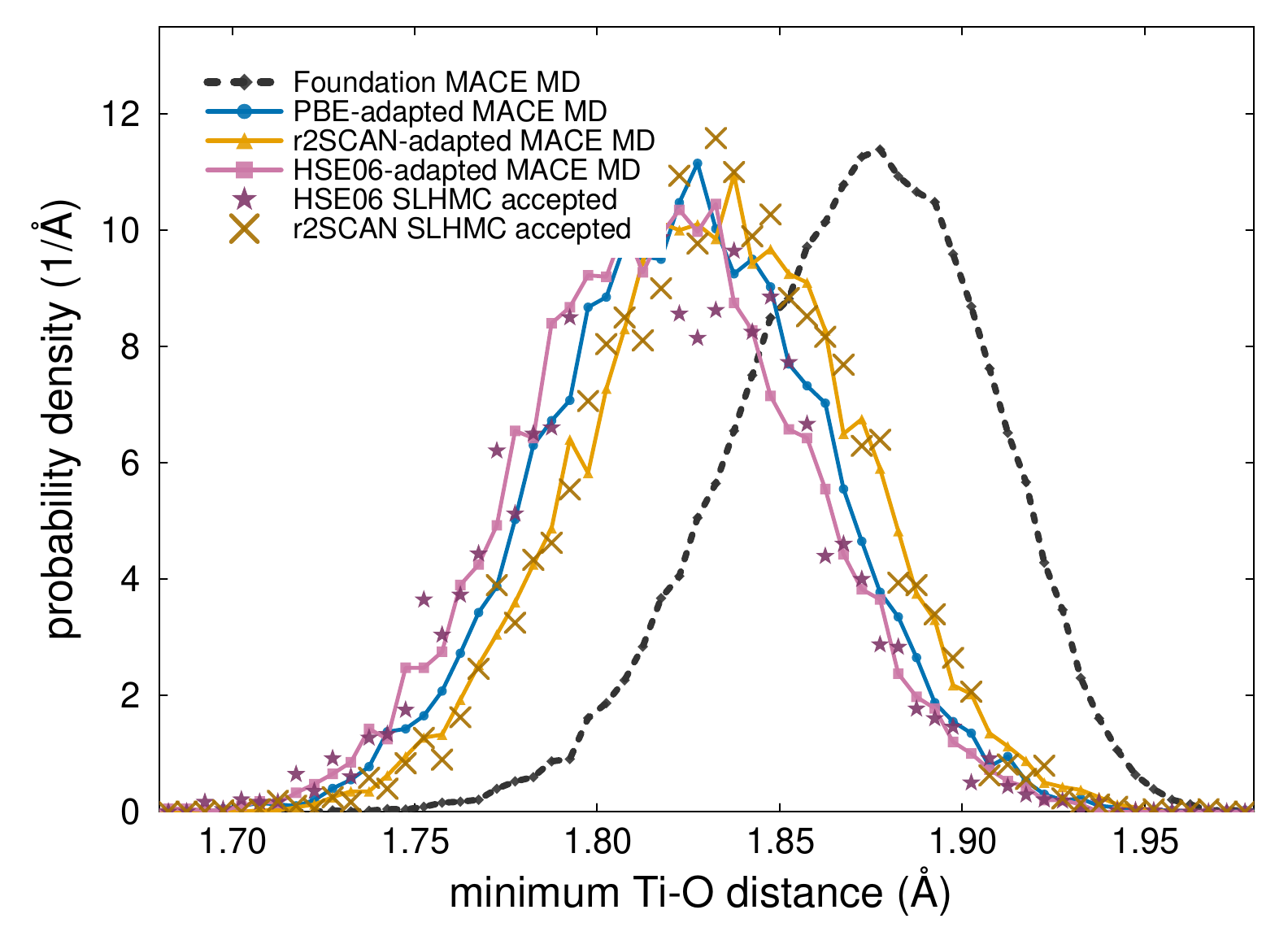}
\caption{Distribution of nearest-neighbor Ti--O distances across different sampling and fine-tuning strategies. The probability density of the minimum Ti--O distance is obtained from MACE-driven molecular dynamics or from accepted SLHMC states.}
\label{fig:nearest-tio}
\end{figure}

Figure~\ref{fig:nearest-tio} shows that the foundation model exhibits a systematic shift relative to the target-functional distributions, indicating that the pretrained potential samples a different local ensemble. The mean minimum Ti--O distance is 1.871 \AA{} for the foundation-model trajectory, compared with 1.824 \AA{} for the PBE-adapted MACE trajectory, 1.833 \AA{} for the r$^2$SCAN-adapted trajectory, and 1.818 \AA{} for the HSE06-adapted trajectory. Block-bootstrap standard errors for these means are below 0.001 \AA{} for all datasets considered here, so the tens-of-m\AA{} shifts are well resolved relative to the finite trajectory uncertainty.

After SLHMC-based fine-tuning, the nearest-neighbor distribution moves toward that obtained from target-functional sampling. This behavior is most relevant for HSE06, where direct ab initio MD sampling is computationally demanding and therefore rarely used for data generation. The mean minimum Ti--O distance from HSE06-adapted MACE MD is 1.8178 \AA{}, close to the HSE06 SLHMC accepted-state value of 1.8173 \AA{}. For r$^2$SCAN, the corresponding values are 1.8332 and 1.8329 \AA{}. In both cases, the differences between the fine-tuned MACE trajectory and the target-anchored reference are smaller than the corresponding bootstrap standard errors of about 0.0005--0.0007 \AA{}, indicating that the fine-tuned model captures not only local interactions but also the statistical weight of relevant configurations within the present finite-sample resolution.

The same conclusion is obtained from a distribution-level metric. Using the one-dimensional Wasserstein distance \cite{Vallender1974-w1} between nearest-neighbor Ti--O distributions, defined for the present analysis in Appendix Eq.~\ref{eq:w1-empirical}, the mismatch to the HSE06 reference decreases from 0.0538 \AA{} for the foundation model to 0.0009 \AA{} for the HSE06-adapted MACE trajectory. For r$^2$SCAN, the corresponding mismatch decreases from 0.0382 to 0.0008 \AA{}. These values quantify the visual trend in Fig.~\ref{fig:nearest-tio}: the fine-tuned models move much closer to the target-functional reference distributions than the foundation-model control.

These observations show that a key limitation of conventional fine-tuning is not simply insufficient label accuracy, but a mismatch in the sampled configurational distribution. Configurations corresponding to short-bond or compressed environments are underrepresented in datasets generated from lower-level dynamics, leading to systematic bias in the learned potential.

\subsection{Radial distribution functions converge toward the target-functional ensemble}

\begin{figure}[htbp]
\centering
\includegraphics[width=0.96\textwidth]{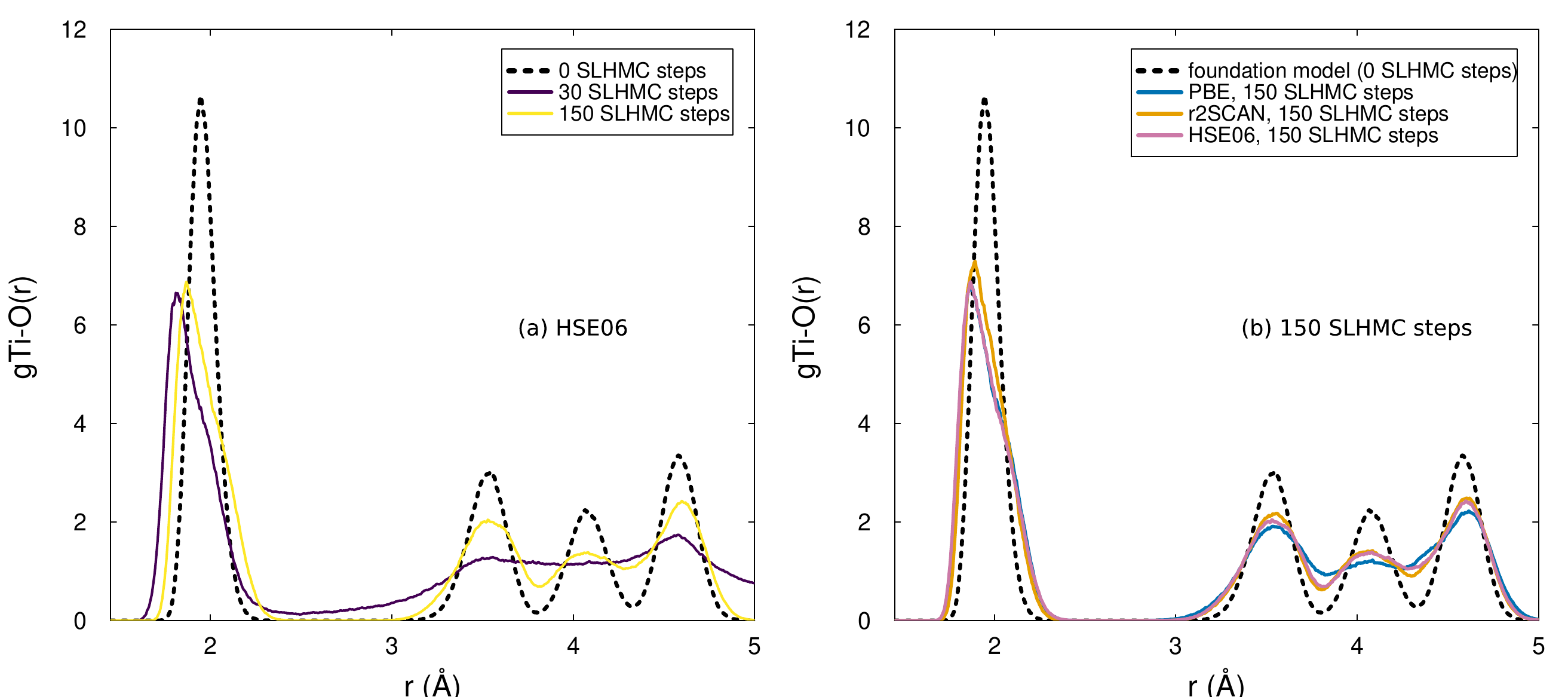}
\caption{Ti--O radial distribution functions during SLHMC fine-tuning. (a) Evolution of $g_{\mathrm{Ti-O}}(r)$ for the HSE06-targeted model at representative checkpoints during sequential fine-tuning. (b) Comparison of PBE, r$^2$SCAN, and HSE06 models after 150 SLHMC steps, together with the foundation model.}
\label{fig:rdf-48}
\end{figure}

We next examine whether the improvement in nearest-neighbor statistics extends to longer-range structural correlations. Figure~\ref{fig:rdf-48} shows the HSE06 case, where distribution-guided sampling is most valuable because direct hybrid-functional MD is the most expensive. The corresponding PBE and r$^2$SCAN RDF evolutions are shown in Appendix Fig.~\ref{fig:rdf-pbe-r2scan}; those panels verify that the same qualitative convergence behavior is not limited to the hybrid-functional target.

Starting from the foundation model, the initial distributions deviate from the target-functional profiles, particularly in the first coordination shell. As additional target-functional evaluations are accumulated through SLHMC and incorporated into the training set, the radial distribution functions move toward the target-functional reference distributions. In the HSE06 comparison, the first-shell peak position and the sixfold Ti--O coordination are recovered by the HSE06-adapted MACE trajectory within the finite-sample uncertainty of the present RDF analysis. Because these comparisons use fixed-cell NVT trajectories, we use them only as structural consistency checks against the SLHMC reference, not as direct comparisons to experiment.

This convergence gives a structural interpretation of the RMSE reduction discussed above. The model is not merely improving pointwise predictions of energies and forces; it is learning a potential energy surface whose equilibrium sampling behavior approaches that of the target functional. The comparison at 150 SLHMC steps is included to show how the functionals separate from the foundation model within the fixed-cell NVT setup. The comparison to experiment is reserved for the NPT cell metrics in Fig.~\ref{fig:npt-cell}.

\begin{figure}[htbp]
\centering
\includegraphics[width=0.90\textwidth]{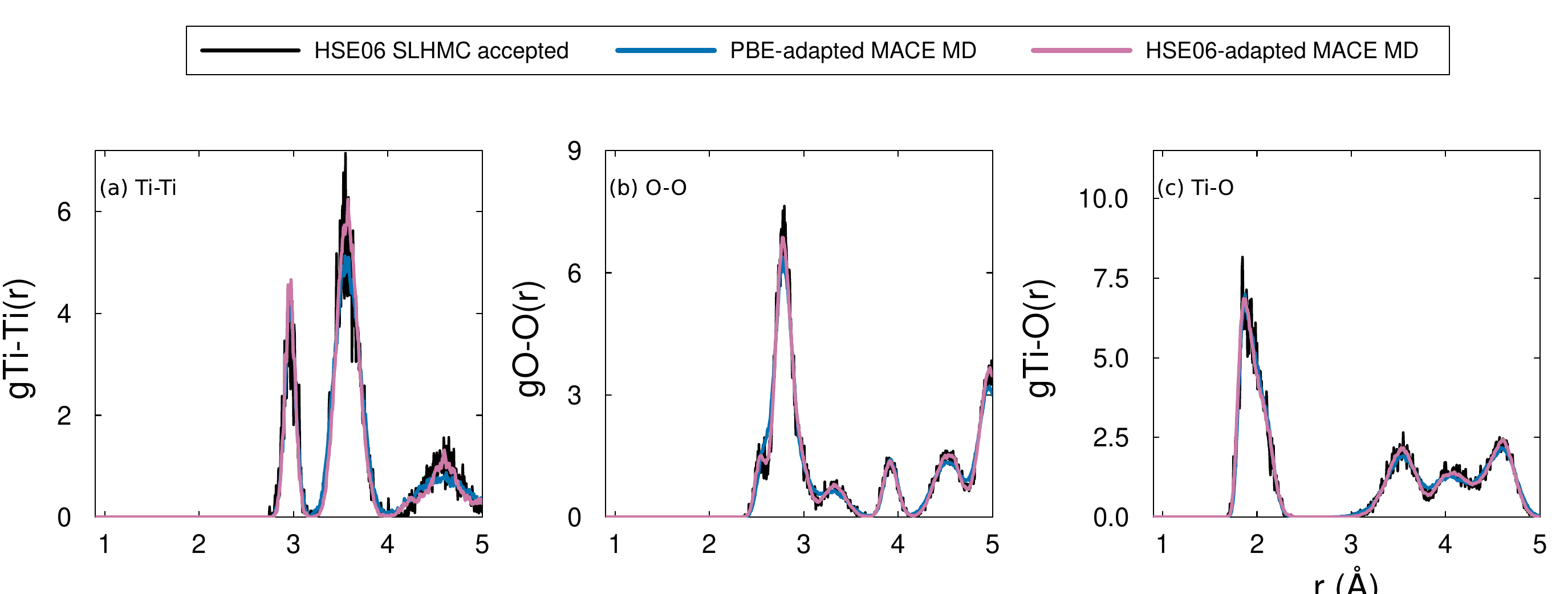}
\caption{Comparison of pair radial distribution functions between SLHMC sampling and MACE-driven molecular dynamics. Radial distribution functions for Ti--Ti, O--O, and Ti--O pairs are obtained from HSE06 SLHMC accepted states and from molecular dynamics using the HSE06-fine-tuned MACE model, with the PBE-adapted trajectory included as a lower-level control.}
\label{fig:rdf-216}
\end{figure}

Figure~\ref{fig:rdf-216} compares pair correlations from SLHMC sampling with those from MLP-driven MD using the HSE06-fine-tuned model. The Ti--Ti, O--O, and Ti--O radial distribution functions agree at the level expected from the finite 48-atom SLHMC reference. For Ti--O, the first peak occurs at 1.883 \AA{} in the HSE06 SLHMC reference and at 1.876 \AA{} in the HSE06-adapted 48-atom MACE MD trajectory, with block-bootstrap uncertainties of about 0.014 and 0.013 \AA{}, respectively, while the integrated coordination remains six in both cases. We therefore use the RDFs as local-structure validation rather than as proof of complete ensemble equivalence.

\subsection{Distribution-consistent fine-tuning improves equilibrium cell properties}

To assess whether the improvements in structural distributions translate into thermodynamic observables, we examine the evolution of equilibrium lattice parameters and cell volume under constant-pressure conditions. These quantities provide a stringent test, as they depend on the free-energy balance over the sampled ensemble rather than on a limited subset of configurations.

\begin{figure}[htbp]
\centering
\includegraphics[width=0.9\textwidth]{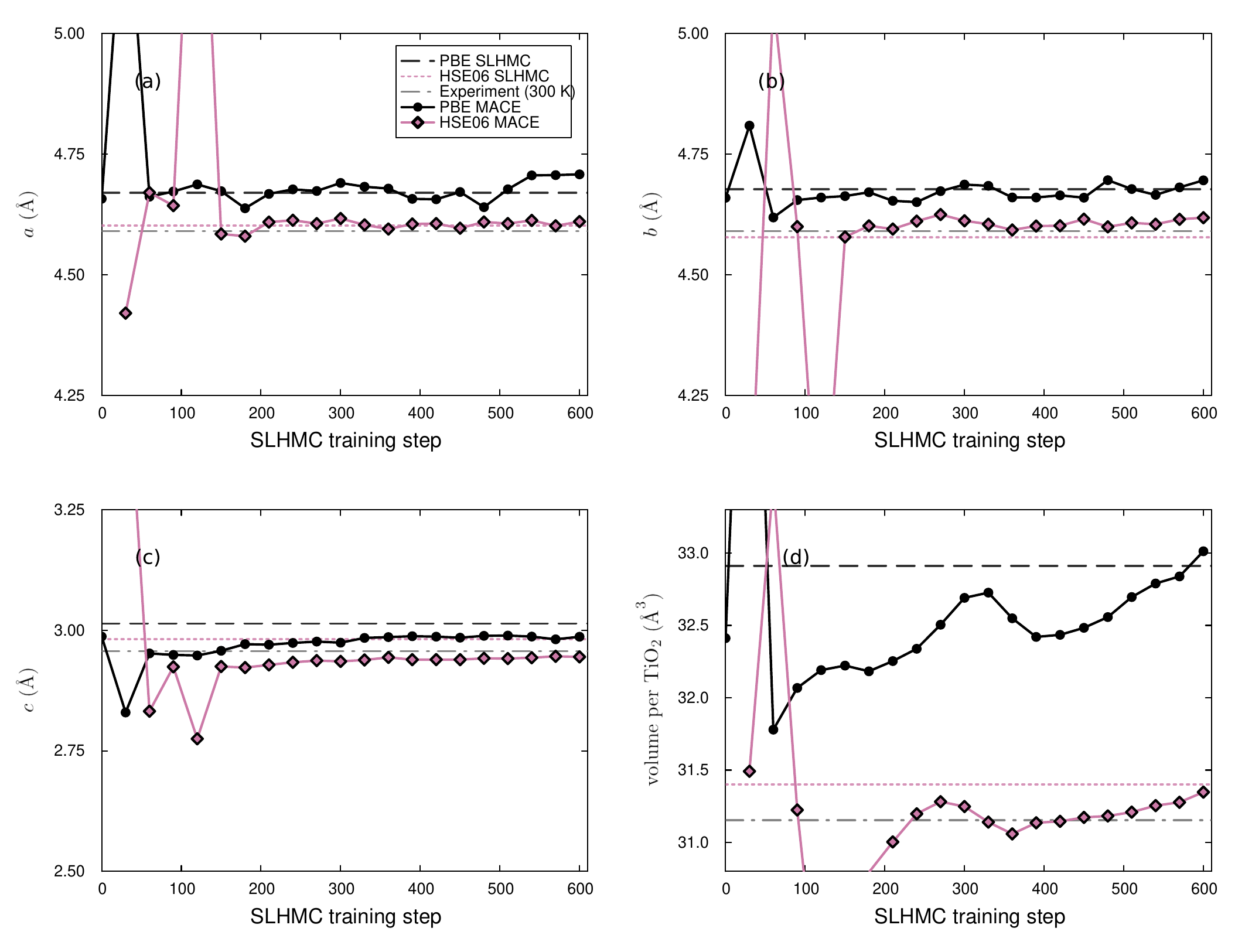}
\caption{Convergence of 300 K NPT lattice parameters and volume during SLHMC-based fine-tuning. Solid curves denote 48-atom MACE-driven NPT trajectories for the PBE and HSE06 target functionals. Horizontal reference lines denote the corresponding SLHMC NPT values. In panels (a)--(c), the lattice constants are reported in the minimal rutile-cell convention. The gray dash-dotted line marks the experimental 300 K rutile reference from Ref.~\cite{Hummer2007-rutile}.}
\label{fig:npt-cell}
\end{figure}

Figure~\ref{fig:npt-cell} compares how the equilibrium structural parameters evolve during SLHMC-based fine-tuning. The production MACE NPT trajectories shown as solid curves were propagated in 48-atom cells at 300 K and ambient pressure (0.101325 MPa). For both PBE and HSE06, these 300 K production trajectories used MACE models fine-tuned on separate 600 K NPT-SLHMC proposal datasets; the 600 K label refers to the temperature of the training data, not to the temperature of the plotted evaluation runs. The corresponding SLHMC NPT reference values come from separate 48-atom runs at the same temperature and pressure as the evaluation trajectories. In Fig.~\ref{fig:npt-cell}(a)--(c), the plotted lattice constants are converted to the minimal rutile-cell convention by dividing the 48-atom $2\times2\times2$ cell lengths by two, allowing direct comparison with the 300 K experimental lattice constants of rutile TiO$_2$ \cite{Hummer2007-rutile}. Panel (d) reports the volume per TiO$_2$ formula unit.

The PBE-targeted model evolves from 32.41 \AA$^3$ per TiO$_2$ for the foundation-model control toward the PBE SLHMC reference value of 32.91 \AA$^3$ per TiO$_2$ as additional SLHMC updates are incorporated. The HSE06-targeted model moves in the opposite direction and remains much closer to the experimental 300 K volume of 31.15 \AA$^3$ per TiO$_2$. At the final analyzed point, the estimated standard errors from cell fluctuations are about 0.07 and 0.06 \AA$^3$ per TiO$_2$ for the PBE MACE and PBE SLHMC series, respectively. For HSE06, the corresponding values are about 0.06 and 0.08 \AA$^3$ per TiO$_2$.

As an auxiliary control, we also trained an HSE06 MACE model on 600 HSE06-labeled configurations drawn from a 600 K foundation-model NPT trajectory and then propagated that relabeled model in a separate 300 K NPT run. Its final volume is 32.27 \AA$^3$ per TiO$_2$, well above both the HSE06 SLHMC NPT reference (31.40 \AA$^3$ per TiO$_2$) and the target-guided HSE06 MACE NPT result (31.35 \AA$^3$ per TiO$_2$). This control shows that, in the NPT setting, where the target labels are placed matters in addition to their fidelity.

Thus, the effect of distribution mismatch is visible not only in local structural descriptors but also in macroscopic observables derived from ensemble averages. The HSE06 trend remains somewhat noisier than the PBE counterpart even over the same analyzed step range, indicating that the hybrid-functional target remains the more demanding transfer problem under the present workflow.

The experimental rutile volume at 300 K is approximately 31.15 \AA$^3$ per TiO$_2$ \cite{Hummer2007-rutile}. Within the present 48-atom workflow, the HSE06-targeted NPT values are closer to this experimental reference than the corresponding PBE values. The lattice constants and volume in Fig.~\ref{fig:npt-cell} are taken directly from the 300 K diffraction data reported by Hummer \textit{et al.}\ \cite{Hummer2007-rutile}. A complementary comparison based on the crystallographic short/long Ti--O first-shell decomposition derived from Howard \textit{et al.}\ \cite{Howard1991-rutile} is summarized in Appendix Table~\ref{tab:npt-first-shell}. That comparison points in the same direction: within the present setup, the HSE06 SLHMC NPT reference is more consistent with the experimental rutile first-shell geometry than the PBE SLHMC NPT reference.

\begin{table*}[htbp]
\caption{Summary of key structural and cell metrics discussed in the main text. The table is separated into an NVT structural block and an NPT cell-metric block so that the underlying ensembles remain explicit. Reported uncertainties are standard errors estimated from the finite trajectories as described in Methods. The NVT block summarizes fixed-cell trajectories: the foundation control uses a 216-atom $3\times3\times4$ rutile supercell, whereas the other NVT entries use a 48-atom $2\times2\times2$ rutile supercell with the same minimal-cell parameters. The experimental rutile volume at 300 K is taken from Ref.~\cite{Hummer2007-rutile}. In the NPT block, the reported volume quantity is $V/\mathrm{TiO_2}$, matching panel (d) of Fig.~\ref{fig:npt-cell}.}
\label{tab:key-metrics}
\centering
\footnotesize
\resizebox{\textwidth}{!}{%
\begin{tabular}{@{}llccc@{}}
\toprule
Model / sampling & Ensemble & Mean min Ti--O (\AA) & Ti--O first peak (\AA) & Ti--O coordination \\
\midrule
Foundation MACE control & NVT, 300 K & $1.8711 \pm 0.0004$ & $1.9491 \pm 0.0056$ & $6.0000 \pm 0.0000$ \\
PBE-adapted MACE MD & NVT, 300 K & $1.8242 \pm 0.0006$ & $1.8736 \pm 0.0122$ & $6.0000 \pm 0.0000$ \\
r$^2$SCAN-adapted MACE MD & NVT, 300 K & $1.8332 \pm 0.0007$ & $1.8874 \pm 0.0145$ & $6.0000 \pm 0.0000$ \\
HSE06-adapted MACE MD & NVT, 300 K & $1.8178 \pm 0.0007$ & $1.8757 \pm 0.0128$ & $5.9994 \pm 0.0003$ \\
r$^2$SCAN SLHMC accepted-state reference & NVT, 300 K & $1.8329 \pm 0.0005$ & $1.8949 \pm 0.0233$ & $5.9999 \pm 0.0001$ \\
HSE06 SLHMC accepted-state reference & NVT, 300 K & $1.8173 \pm 0.0007$ & $1.8832 \pm 0.0143$ & $5.9999 \pm 0.0001$ \\
\midrule
\multicolumn{5}{@{}l@{}}{\textbf{NPT cell metrics at 300 K and ambient pressure (0.101325 MPa)}} \\
\midrule
\multicolumn{2}{@{}l}{Model / sampling} & \multicolumn{2}{c}{Volume metrics} & \\
\cmidrule(lr){3-4}
\multicolumn{2}{@{}l}{} & $V/\mathrm{TiO_2}$ (\AA$^3$) & $\Delta V_{\mathrm{exp}}$ (\AA$^3$) & \\
\midrule
PBE-adapted MACE MD & NPT, ambient pressure & $33.0123 \pm 0.0678$ & $+1.8594$ & \\
PBE SLHMC NPT reference & NPT, ambient pressure & $32.9125 \pm 0.0552$ & $+1.7596$ & \\
HSE06-adapted MACE MD & NPT, ambient pressure & $31.3476 \pm 0.0578$ & $+0.1947$ & \\
HSE06 SLHMC NPT reference & NPT, ambient pressure & $31.4015 \pm 0.0833$ & $+0.2486$ & \\
HSE06 relabeled foundation-NPT control & NPT, ambient pressure & $32.2657 \pm 0.0550$ & $+1.1128$ & \\
Experiment \cite{Hummer2007-rutile} & NPT, ambient pressure & $31.1529$ & $0.0000$ & \\
\bottomrule
\end{tabular}
}
\end{table*}

Table~\ref{tab:key-metrics} collects the main structural and NPT metrics while keeping the NVT and NPT ensembles separate. The fine-tuned MACE trajectories move away from the foundation-model controls and toward the corresponding SLHMC reference values. For the HSE06 and r$^2$SCAN structural metrics in NVT, the remaining differences are small on the scale of the finite-trajectory uncertainties. In the NPT block, the main comparison is the cell volume. The experimental row and the $\Delta V_{\mathrm{exp}}$ column provide an external check that the HSE06 NPT values lie closer to the 300 K rutile volume than the corresponding PBE values, and that the off-target HSE06 relabeling control remains substantially farther from both the HSE06 SLHMC reference and experiment.

\section{Discussion}\label{sec12}

The results show that, in cross-functional fine-tuning of MLIPs, the sampling distribution is part of the learning problem. Forward RMSE values measure local prediction accuracy, whereas the structural analyses test whether the learned potential samples the same regions of configuration space as the target functional.

The accepted SLHMC trajectory and the supervised training dataset are not identical. For a fixed proposal potential, the accepted chain targets the Boltzmann distribution of the DFT functional used in the Metropolis test. The supervised dataset, by contrast, consists of target-functional evaluations along the proposal stream and includes rejected configurations. Because the proposal model is updated during self-learning, we interpret the full stream as target-guided data generation rather than as a single equilibrium trajectory.

The nearest-neighbor and radial-distribution analyses show why this distinction matters. Conventional lower-level sampling biases the dataset away from configurations favored by the target functional. Once those configurations are absent or severely underrepresented, they are difficult to recover by fine-tuning alone, even if the labels themselves are accurate. Our goal here is to isolate that sampling effect, not to provide a full ranking of alternative active-learning or data-selection strategies at fixed label budget.

The NPT cell parameters show that the consequences extend beyond local structure to macroscopic observables. In the present TiO$_2$ example, the HSE06-targeted NPT values lie closer to the experimental 300 K rutile volume than the corresponding PBE values, consistent with the known tendency of semilocal GGAs to overestimate rutile equilibrium volumes. The auxiliary HSE06 relabeling control strengthens the same point: despite using the same target functional for labeling, an off-target 600 K foundation-model trajectory yields a final volume of 32.27 \AA$^3$ per TiO$_2$, far above both the HSE06 SLHMC NPT reference and the target-guided HSE06 MACE NPT result.

The present NPT results are nevertheless not fully converged. Even the HSE06-adapted MACE NPT trajectory does not yet coincide exactly with the HSE06 SLHMC reference, and the local Ti--O structure remains harder to reproduce in NPT than in NVT. We therefore view the remaining mismatch primarily as a data-coverage problem: under the present workflow, NPT fine-tuning is more demanding and would likely benefit from longer NPT SLHMC sampling and a larger accumulated training set.

Although the present study focuses on TiO$_2$, the underlying mechanism is not system specific. Any cross-functional fine-tuning problem in which the target functional alters the statistical weight of configurations is expected to be affected by similar distribution mismatch. This includes transfer across different density functionals, correction schemes, or levels of electronic correlation.

The main limitation of the current work is that the analysis is restricted to a single material system and a finite set of thermodynamic conditions. Future work should test the approach across a broader range of materials, phases, and sampling regimes. The present results nevertheless show that accurate target-level labels alone are insufficient when the configurational distribution is incorrect.

In summary, we have shown that a key limitation of conventional cross-functional fine-tuning workflows is not only the availability of high-fidelity labels, but also the mismatch between the sampling distribution and the target Boltzmann distribution. By coupling data generation to target-functional acceptance through SLHMC, foundation MLIPs can be adapted across multiple density functionals, including computationally demanding hybrid functionals. The resulting potentials reproduce target-anchored structural references and the thermodynamic observables examined here more accurately than lower-level and foundation-model sampling controls.

\section{Methods}\label{sec:methods}

\subsection{First-principles calculations for SLHMC}

The simulations were based on the rutile phase of TiO$_2$. The conventional rutile unit cell contains two TiO$_2$ formula units, i.e., six atoms. The target-functional energy and force evaluations used in the SLHMC acceptance tests were carried out for 48-atom supercells, corresponding to a $2 \times 2 \times 2$ replication of this conventional cell. Some machine-learning-potential simulations used larger 216-atom supercells, corresponding to a $3 \times 3 \times 4$ replication of the same rutile conventional cell. The first-principles reference calculations were performed with the GPU-enabled version of VASP 6.5.1 using the projector augmented-wave method, with the PAW\_PBE Ti\_pv potential, which treats the Ti semicore $p$ states as valence, and the PAW\_PBE O potential. The VASP interface used in the present SLHMC workflow employed the patch distributed with the present PIMD release. A plane-wave cutoff energy of 520 eV and Gamma-point-only Brillouin-zone sampling were used for all SLHMC reference calculations. The calculations were spin unpolarized, used Gaussian electronic smearing with a width of 0.05 eV, and imposed no symmetry constraints. Electronic self-consistency was converged to $10^{-4}$ eV. The VASP calls used for SLHMC labeling were single-configuration energy and force evaluations of the proposed structures.

Three exchange--correlation levels were considered in the present manuscript: PBE, r$^2$SCAN, and HSE06. PBE was used as a generalized-gradient approximation, r$^2$SCAN as a meta-GGA functional, and HSE06 as a screened hybrid functional with 25\% short-range exact exchange and a screening parameter of 0.2 \AA$^{-1}$. The same supercell, PAW potentials, plane-wave cutoff, k-point sampling, spin treatment, smearing, and electronic convergence criterion were used for all three targets so that differences between SLHMC datasets arise from the exchange--correlation functional rather than from changes in the numerical setup. The target of the present cross-functional transfer is therefore a consistent target-labeling protocol for each functional, not a claim of fully converged absolute benchmarking at every electronic-structure level.

\subsection{SLHMC and molecular-dynamics sampling}

SLHMC and MLP-driven MD simulations were performed using PIMD version 2.7.2 with the PIMD--MACE interface \cite{Nagai2020-gx,Thomsen2024-dn}. PIMD is a path-integral molecular dynamics package, but all calculations here used one bead, corresponding to the classical-nuclei limit. In the present workflow, PIMD served as the simulation driver for both hybrid Monte Carlo and conventional molecular dynamics, interfacing trajectory propagation to MACE and the acceptance test to VASP 6 target-functional evaluations through its dual-potential implementation.

In SLHMC, trial trajectories were generated with the current MACE potential using a 1 fs integration time step. The endpoint of each trial trajectory was accepted or rejected by a Metropolis test using the target-functional DFT energy, and both accepted and rejected DFT-labeled configurations were retained for subsequent MACE fine-tuning. Rejected endpoints are not samples from the accepted-state reference chain, but they still provide useful information about regions proposed by the current model and penalized by the target functional. Retaining them helps the updated MACE potential avoid repeatedly proposing the same off-target regions in later SLHMC cycles. For a fixed proposal model, the NVT acceptance probability has the standard HMC form
\begin{align}
    p_{\mathrm{acc}}
    =
    \min\left[1,\exp\left(-\beta \Delta H_{\mathrm{target}}\right)\right],
\end{align}
where $\Delta H_{\mathrm{target}}$ is evaluated with the target-functional potential energy at the HMC endpoints and the kinetic-energy terms associated with the sampled momenta. The NPT implementation uses the corresponding extended-ensemble Metropolis test implemented in the PIMD dual-potential driver, following the isothermal--isobaric SLHMC formulation of Kobayashi \textit{et al.}\ \cite{Kobayashi2021-npt-slhmc}. Because the proposal model is updated during self-learning, accepted configurations from the adaptive stage are target anchored but are not treated as a single long production trajectory from a frozen kernel.

The NVT SLHMC simulations used 48 atoms at 300 K. All 48-atom NVT structural analyses in the main text used a fixed $2\times2\times2$ rutile conventional cell with $a=b=9.1997$ \AA{} and $c=5.9184$ \AA{}, corresponding to $a=b=4.5998$ \AA{} and $c=2.9592$ \AA{} in the minimal rutile-cell convention. The 216-atom foundation control used the corresponding fixed $3\times3\times4$ replication of the same rutile cell.

For the adaptive-HMC summary, the analyzed runs contain 600 SLHMC cycles for PBE, 600 for r$^2$SCAN, and 600 for HSE06. For the forward-error comparison in Fig.~\ref{fig:forward-rmse}, all three functionals are plotted only through 390 steps so that the prequential comparison window is identical across PBE, r$^2$SCAN, and HSE06 and remains focused on the early-to-intermediate stage of NVT fine-tuning. The number of MACE MD steps used to generate one HMC proposal was not fixed throughout a run. Instead, we started from a short proposal trajectory and adjusted the proposal length every 30 SLHMC cycles. The resulting cumulative acceptance ratios and adapted proposal lengths are shown in Appendix Fig.~\ref{fig:hmc-adaptation}. The accepted-state trajectories used as structural references in the nearest-neighbor and pair-RDF analyses were analyzed separately; the HSE06 and r$^2$SCAN accepted-state datasets used there each contain 600 frames.

The NPT references used in Fig.~\ref{fig:npt-cell} were likewise analyzed separately for each functional: both the PBE and HSE06 SLHMC references come from completed 48-atom NPT runs at 300 K and ambient pressure (0.101325 MPa). The production MACE NPT simulations used for the main-text cell-parameter comparison were then propagated separately in 48-atom cells at the same temperature and pressure. For PBE and HSE06, however, the underlying MACE models were fine-tuned on 600 K NPT-SLHMC proposal data before being evaluated in these separate 300 K NPT runs. The corresponding production MACE NPT series shown for PBE and HSE06 each extend to 600 fine-tuning steps. As an auxiliary off-target control for the NPT discussion, we also used a separate 48-atom foundation-model NPT trajectory at 600 K and ambient pressure (0.101325 MPa), propagated for 130000 MD steps and sampled every 100 steps to generate a pool of 1200 candidate configurations. From this pool, 600 configurations were evaluated at the HSE06 level and used to fine-tune a separate HSE06 MACE model, which was then propagated in a separate 48-atom NPT run at 300 K and ambient pressure (0.101325 MPa).

MLP-driven production MD simulations used the same one-bead classical limit. NVT simulations used 48- or 216-atom cells at 300 K, depending on the comparison, and NPT simulations used 48-atom cells at 300 K and ambient pressure (0.101325 MPa) for the main-text cell-parameter comparison. These production trajectories were propagated for 5000 steps using massive Nos\'e--Hoover chain thermostats and the corresponding foundation or fine-tuned MACE potentials. The production trajectories analyzed in the manuscript were sampled every 10 MD steps, giving 500 recorded frames from each 5000-step run. This setting applies only to the production trajectories analyzed in the manuscript, not to the separate 600 K foundation-model NPT trajectory used to build the off-target HSE06 relabeling control. The 48-atom simulations were used for direct comparison with 48-atom SLHMC sampling, whereas the 216-atom simulations were used where larger-cell structural observables were specifically examined. Appendix Table~\ref{tab:protocols} distinguishes these production trajectories from the SLHMC proposal stream used for training and from the accepted-state references used for structural validation.

\subsection{MACE initialization and sequential fine-tuning}

The initial machine-learning interatomic potential was the small MACE-MP-0 foundation model \cite{NEURIPS2022_4a36c3c5,Batatia2025-vv}. We used the small MACE-MP-0 checkpoint from the public release, and all MACE training and inference calculations in this workflow used MACE version 0.3.14. MACE-MP-0 was pretrained on the Materials Project trajectory dataset (MPTrj), which contains static calculations and structural relaxation trajectories for a broad range of inorganic materials at the PBE/GGA and GGA+$U$ levels of theory. The small model is a ScaleShiftMACE architecture with 128 scalar hidden channels, two interaction layers, correlation order 3, spherical harmonics up to $\ell = 3$, and a radial cutoff of 6.0 \AA. The original 89-element foundation model contains 3,847,696 trainable parameters.

During SLHMC, this MACE model was fine-tuned sequentially using the DFT-labeled configurations accumulated up to each update point. All trainable parameters of the small MACE-MP-0 model were updated during fine-tuning; no layers were frozen. Fine-tuning was performed every 30 SLHMC steps, after which the updated model was used as the active proposal potential for the next sampling cycle. At each fine-tuning cycle, 10\% of the accumulated configurations were randomly assigned to validation and the remaining configurations were used for training; no independent test set was used during training. Model snapshots were saved after each cycle.

Fine-tuning was carried out using the Adam optimizer with a learning rate of 0.01 and a weight decay of $5 \times 10^{-7}$. All MACE training and inference calculations in the SLHMC workflow were performed in double precision (float64) on a single NVIDIA H100 GPU. The maximum number of epochs was 50, with early-stopping patience of 10 epochs and learning-rate scheduler patience of 5 epochs. The training and validation batch sizes were both 8. The loss function was a weighted energy--force loss with weights of 1.0 for energies and 100.0 for forces. Isolated atomic energies were estimated using the average-energy option.

\subsection{Error and structural analyses}

Forward prediction errors were evaluated in a prequential manner. For a MACE model trained using configurations accumulated through SLHMC step $t$, predictions were evaluated only on the next 30 DFT-labeled configurations, from steps $t+1$ to $t+30$, which were not included in the training data for that model. Energy errors were reported as RMSE values in meV per atom, and force errors were reported as RMSE values in meV \AA$^{-1}$ over all Cartesian force components. These prequential errors are intended to monitor forward prediction during adaptive sampling, not to define an independent benchmark of global model accuracy.

Nearest-neighbor distributions were computed from the minimum Ti--O distance for each Ti atom in each sampled configuration using periodic minimum-image distances. Histograms were normalized as probability densities using a bin width of 0.005 \AA. For the mean nearest-neighbor distance reported in the text, we first averaged the minimum Ti--O distances over Ti atoms within each recorded frame and then estimated the uncertainty of the trajectory mean by block bootstrap over contiguous 10-frame blocks with 400 resampled trajectories.

Radial distribution functions were computed separately for Ti--Ti, O--O, and Ti--O pairs from the periodic simulation cells; the RDF grid spacing was 0.01 bohr, corresponding to 0.00529 \AA. The recorded trajectory frames were analyzed without additional reweighting. The same analysis procedure was applied to SLHMC configurations and to MLP-driven MD trajectories so that differences in the reported distributions reflect the sampling and fine-tuning protocol rather than changes in the post-processing definition. For the first-shell peak positions quoted for the pair RDFs, we reconstructed the peak location from per-frame pair-distance lists by forming a histogram over the corresponding search window, identifying the maximum bin, and applying a local quadratic interpolation through that bin and its two nearest neighbors. Uncertainties for these peak positions were then estimated by block bootstrap over contiguous 10-frame blocks with 300 resampled trajectories. The peak search windows were 3.0--4.0 \AA{} for Ti--Ti, 2.3--3.2 \AA{} for O--O, and 1.6--2.2 \AA{} for Ti--O. The Ti--O coordination number was computed from the number of O atoms within 2.4 \AA{} of each Ti atom and averaged over Ti atoms in each frame; its uncertainty was estimated by the same 10-frame block-bootstrap procedure with 400 resampled trajectories.

For the NPT cell metrics, the mean box matrix and the recorded box fluctuations were read from the simulation outputs. Standard errors for $a$, $b$, and $c$ were taken from the corresponding diagonal cell fluctuations, and the uncertainty of the cell volume was obtained by first-order propagation of these fluctuations for the nearly diagonal simulation cells used here. These propagated standard errors are reported only as measures of finite-trajectory uncertainty within the present simulations; they do not account for systematic effects associated with thermostatting, barostatting, finite size, or the finite length of the SLHMC NPT reference trajectories.

\backmatter


\bmhead{Acknowledgements}
The work of Y.N. was partially supported by JSPS KAKENHI Grant Numbers 22H05114, 26K00651 and 26K01464.
M.S. acknowledges support from JSPS KAKENHI Grant Numbers 24K01145, 25K01629, and 26K01464. 
B.T. acknowledges support from JSPS KAKENHI Grant Numbers 24K01408 and 26K01464. 
This work was partially supported by ``Joint Usage/Research Center for Interdisciplinary Large-scale Information Infrastructures (JHPCN)'' in Japan (Project ID: jh250046).

\section*{Declarations}

\bmhead{Competing interests}
The authors declare no competing interests.

\bmhead{Author contributions}
Y.N. conceived the study, performed the SLHMC and MACE calculations, analyzed the data, and wrote the manuscript. B.T. and M.S. contributed to the PIMD--MACE/SLHMC methodology, interpretation of the results, and manuscript revision. All authors reviewed and approved the manuscript.

\bmhead{Data and code availability}
The simulations reported here used PIMD version 2.7.2 together with the PIMD--MACE interface and the VASP patch distributed with that release, VASP 6.5.1, and MACE version 0.3.14, as described in the Methods. 
The source data used to generate the figures, along with the analysis scripts and workflow input files needed to reproduce the manuscript results, will be made publicly available in a GitHub repository. The repository URL will be added here after the public release of the dataset and code package.

\bmhead{Funding}
The funding sources are listed in the Acknowledgements.

\begin{appendices}
\renewcommand{\thefigure}{A\arabic{figure}}
\renewcommand{\theHfigure}{appendix.A\arabic{figure}}
\renewcommand{\thetable}{A\arabic{table}}
\renewcommand{\theHtable}{appendix.A\arabic{table}}
\setcounter{figure}{0}
\setcounter{table}{0}

\section{Protocol definitions and additional RDF evolution}

This Appendix collects supporting material that is referred to from the main text but would interrupt the narrative if kept in the main Results or Methods sections. Table~\ref{tab:protocols} defines the trajectory labels used throughout the structural comparisons. The figures that follow provide two complementary checks: additional Ti--O RDF evolution for the semilocal targets and the adaptive behavior of the HMC proposal length during SLHMC.

\begin{table}[htbp]
\caption{Trajectory labels used in the structural comparisons. ``MACE MD'' denotes ordinary molecular dynamics with a frozen MACE potential, whereas ``SLHMC accepted states'' denotes configurations accepted by a Metropolis test using target-functional DFT energies. The SLHMC proposal stream, including both accepted and rejected endpoints, is the dataset used for sequential fine-tuning.}
\label{tab:protocols}
\centering
\footnotesize
\renewcommand{\arraystretch}{1.18}
\begin{tabular}{@{}p{0.18\textwidth}p{0.25\textwidth}p{0.23\textwidth}p{0.20\textwidth}@{}}
\toprule
Label in text/figures & How configurations were generated & Use of target DFT & Main use in this work \\
\midrule
Foundation MACE MD & MACE-MP-0 molecular dynamics & none during MD & pretrained-model baseline \\
PBE-, r$^2$SCAN-, or HSE06-adapted MACE MD & molecular dynamics with a frozen fine-tuned MACE model & target DFT used earlier during SLHMC training & predicted target-functional ensemble \\
PBE-adapted MACE control & molecular dynamics with a lower-level or differently fine-tuned MACE model & none during MD & distribution-mismatch control \\
r$^2$SCAN or HSE06 SLHMC accepted & accepted states from an SLHMC Markov chain & DFT energies used in Metropolis test & target-anchored structural reference \\
SLHMC proposal stream & endpoints of MACE HMC proposals, accepted and rejected & DFT energies and forces evaluated for each endpoint & sequential fine-tuning data \\
Forward test blocks & next 30 DFT-evaluated SLHMC endpoints after each update & DFT labels used only for testing that model snapshot & prequential RMSE evaluation \\
\bottomrule
\end{tabular}
\renewcommand{\arraystretch}{1}
\end{table}

For the nearest-neighbor analysis in Fig.~\ref{fig:nearest-tio}, we use the one-dimensional Wasserstein-$1$ distance between the empirical distributions of minimum Ti--O distances. If $\hat{F}_x(r)$ and $\hat{F}_y(r)$ denote the empirical cumulative distribution functions built from two finite samples $\{x_i\}_{i=1}^n$ and $\{y_j\}_{j=1}^m$, the reported metric is
\begin{equation}
W_1(\hat{F}_x,\hat{F}_y) = \int_{-\infty}^{\infty} \left| \hat{F}_x(r) - \hat{F}_y(r) \right| \, \mathrm{d}r.
\label{eq:w1-empirical}
\end{equation}
In one dimension this is equivalent to the earth-mover distance between the two empirical distributions and has the same unit as the underlying coordinate, here \AA. In practice, we evaluate Eq.~\ref{eq:w1-empirical} exactly for the empirical samples by sorting the two distance lists and integrating the absolute difference of the piecewise-constant empirical CDFs between successive sample locations.

For the NPT local-structure check discussed briefly in the main text, we compared the Ti--O first shell using the same split-shell convention as the crystallographic analysis of rutile TiO$_2$ \cite{Howard1991-rutile}. In the Appendix table, the experimental short and long Ti--O distances are therefore not taken from the Hummer lattice-constant dataset used in Fig.~\ref{fig:npt-cell}; instead, they are derived from the rutile structural parameters $(a,c,u)$ reported by Howard \textit{et al.}\ \cite{Howard1991-rutile}. In each sampled frame, the six nearest O neighbors of every Ti atom were sorted by distance; the shortest four were averaged as the ``short'' Ti--O distance, the next two were averaged as the ``long'' Ti--O distance, and the mean over all six bonds was also recorded. These per-frame quantities were then averaged over the trajectory, with standard errors estimated by contiguous 10-frame block bootstrap. The resulting values are listed in Table~\ref{tab:npt-first-shell}. This analysis is included as a supplementary local-structure comparison; in the main text we emphasize the NPT cell metrics because they are cleaner and more stable than the corresponding first-shell bond decomposition for the present short 48-atom MACE NPT trajectories.

\begin{table}[htbp]
\caption{Supplementary NPT first-shell Ti--O bond metrics at 300 K and ambient pressure (0.101325 MPa). The experimental row uses the rutile short/long split derived from the neutron-diffraction structural parameters reported by Howard \textit{et al.}\ \cite{Howard1991-rutile}.}
\label{tab:npt-first-shell}
\centering
\footnotesize
\begin{tabular}{@{}lccc@{}}
\toprule
Model / sampling & Short Ti--O (\AA) & Long Ti--O (\AA) & Six-bond mean (\AA) \\
\midrule
PBE-adapted MACE MD & $1.8901 \pm 0.0006$ & $2.1395 \pm 0.0015$ & $1.9732 \pm 0.0002$ \\
PBE SLHMC NPT reference & $1.9123 \pm 0.0024$ & $2.0821 \pm 0.0067$ & $1.9689 \pm 0.0007$ \\
HSE06-adapted MACE MD & $1.8999 \pm 0.0006$ & $2.1151 \pm 0.0013$ & $1.9716 \pm 0.0001$ \\
HSE06 SLHMC NPT reference & $1.9273 \pm 0.0009$ & $2.0376 \pm 0.0022$ & $1.9640 \pm 0.0002$ \\
HSE06 relabeled foundation-NPT control & $1.8893 \pm 0.0007$ & $2.1340 \pm 0.0016$ & $1.9709 \pm 0.0002$ \\
Experiment \cite{Howard1991-rutile} & $1.9486$ & $1.9800$ & $1.9590$ \\
\bottomrule
\end{tabular}
\end{table}

For completeness, we also show the Ti--O RDF evolution for the PBE and r$^2$SCAN target functionals in the same format used in the main-text HSE06 figure. These panels provide the semilocal-functional counterparts to the HSE06 analysis and make it possible to compare how strongly the target functional changes the local Ti--O environment during sequential fine-tuning. In both cases, the largest changes occur in the first-shell peak and the following minimum, showing that the fine-tuned models move away from the broad foundation-model distribution toward function-specific local structure. The PBE evolution is comparatively mild, whereas the r$^2$SCAN panels show a more visible redistribution of first-shell weight, consistent with the intermediate structural shifts discussed in the main text.

\begin{figure}[htbp]
\centering
\includegraphics[width=\textwidth]{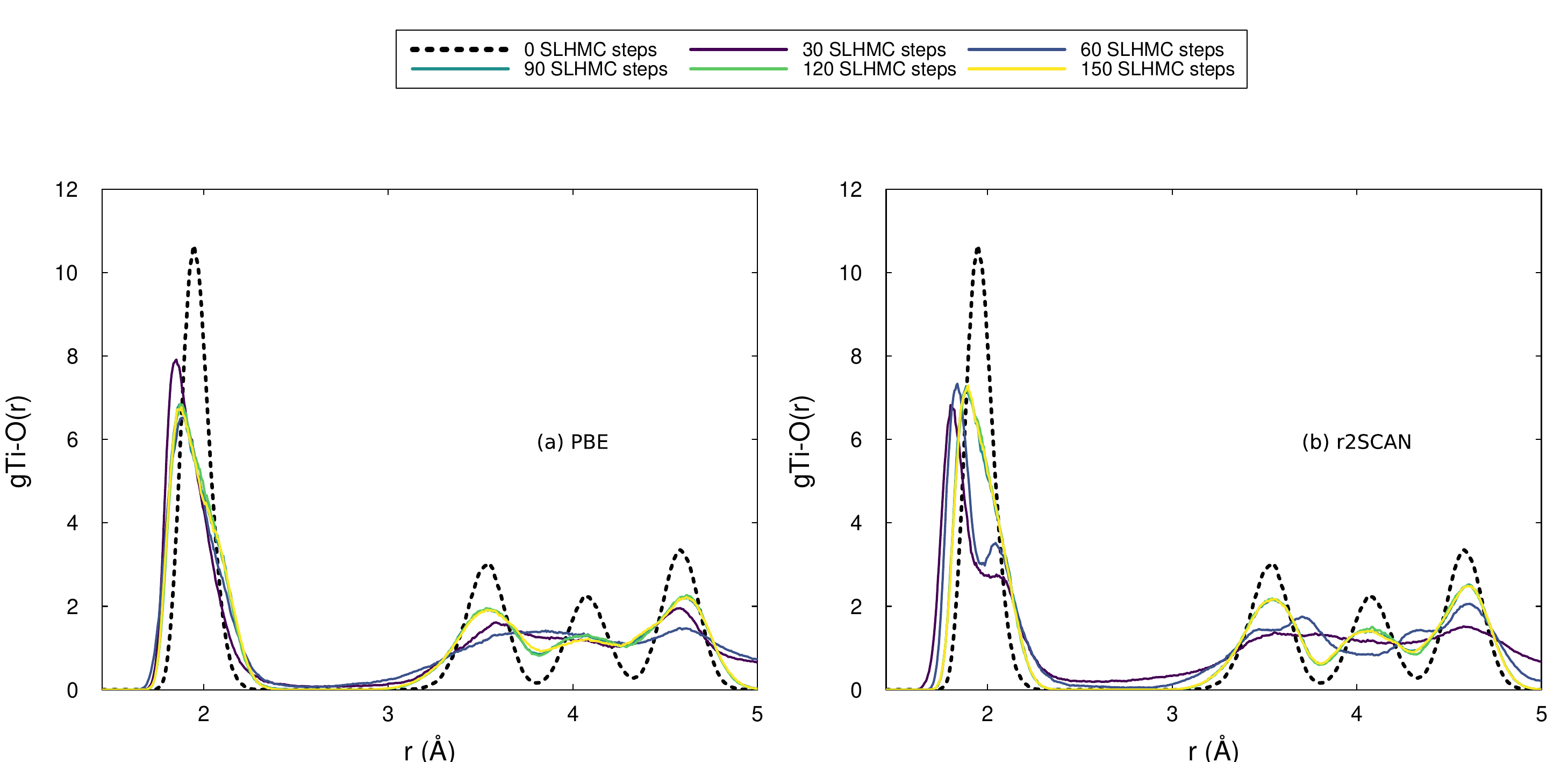}
\caption{Ti--O radial distribution functions during SLHMC fine-tuning for (a) PBE and (b) r$^2$SCAN target functionals. The same sequence of foundation and fine-tuned snapshots is shown for both targets, with the shared legend placed above the panels to keep the first-shell region unobstructed. These panels complement the HSE06-focused RDF evolution shown in Fig.~\ref{fig:rdf-48}.}
\label{fig:rdf-pbe-r2scan}
\end{figure}

\begin{figure}[htbp]
\centering
\includegraphics[width=0.82\textwidth]{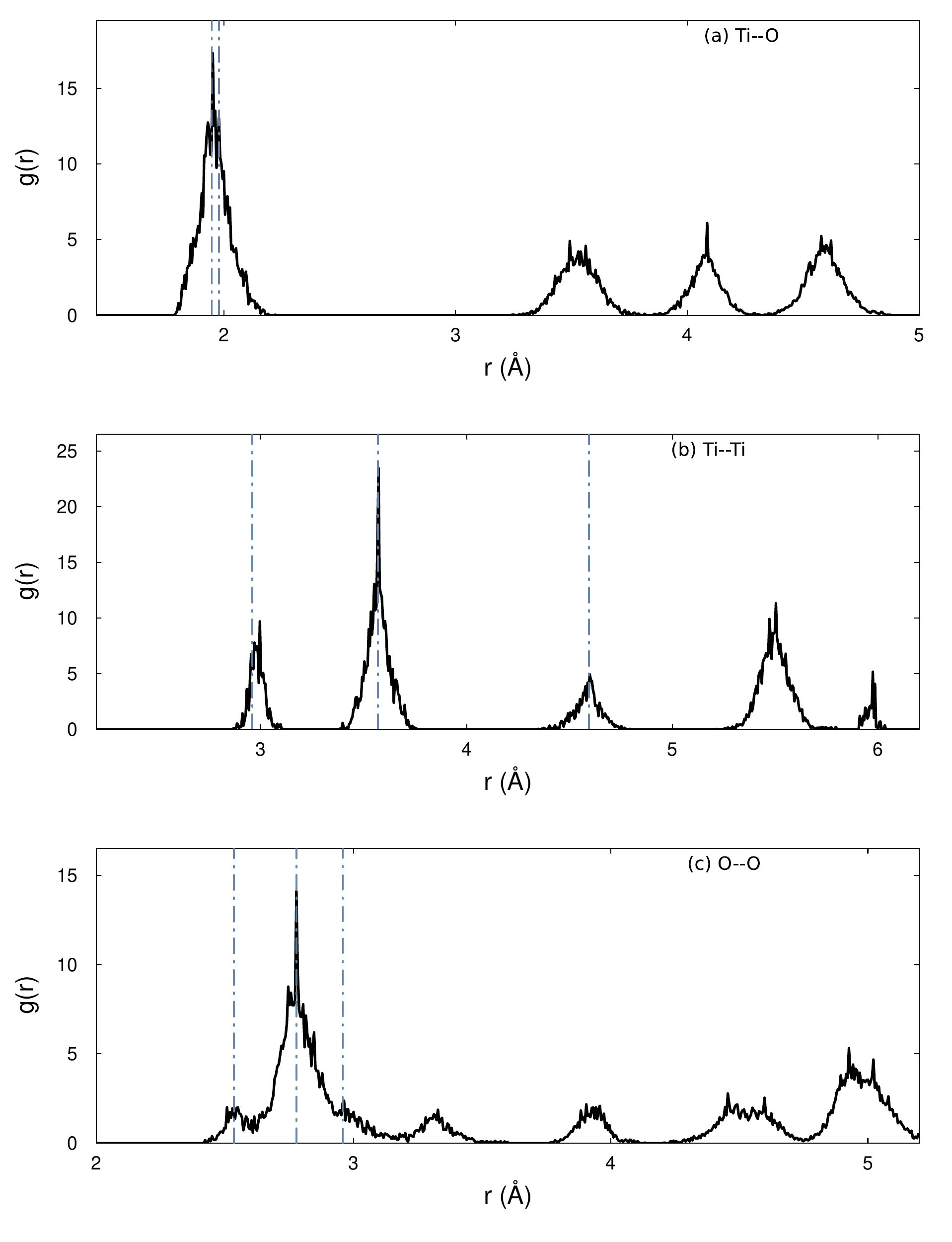}
\caption{Final pair RDFs from the completed HSE06 SLHMC NPT reference trajectory at 300 K and ambient pressure (0.101325 MPa): (a) Ti--O, (b) Ti--Ti, and (c) O--O. Blue dash-dotted guide lines mark representative 300 K rutile distances derived from Ref.~\cite{Howard1991-rutile}.}
\label{fig:hse06-npt-rdf}
\end{figure}

As an additional structural reference, Fig.~\ref{fig:hse06-npt-rdf} shows the final pair RDFs from the completed HSE06 SLHMC NPT run at 300 K and ambient pressure (0.101325 MPa). We include these curves as supplementary HSE06 NPT reference data for rutile TiO$_2$ under the present workflow.

\begin{figure}[htbp]
\centering
\includegraphics[width=0.88\textwidth]{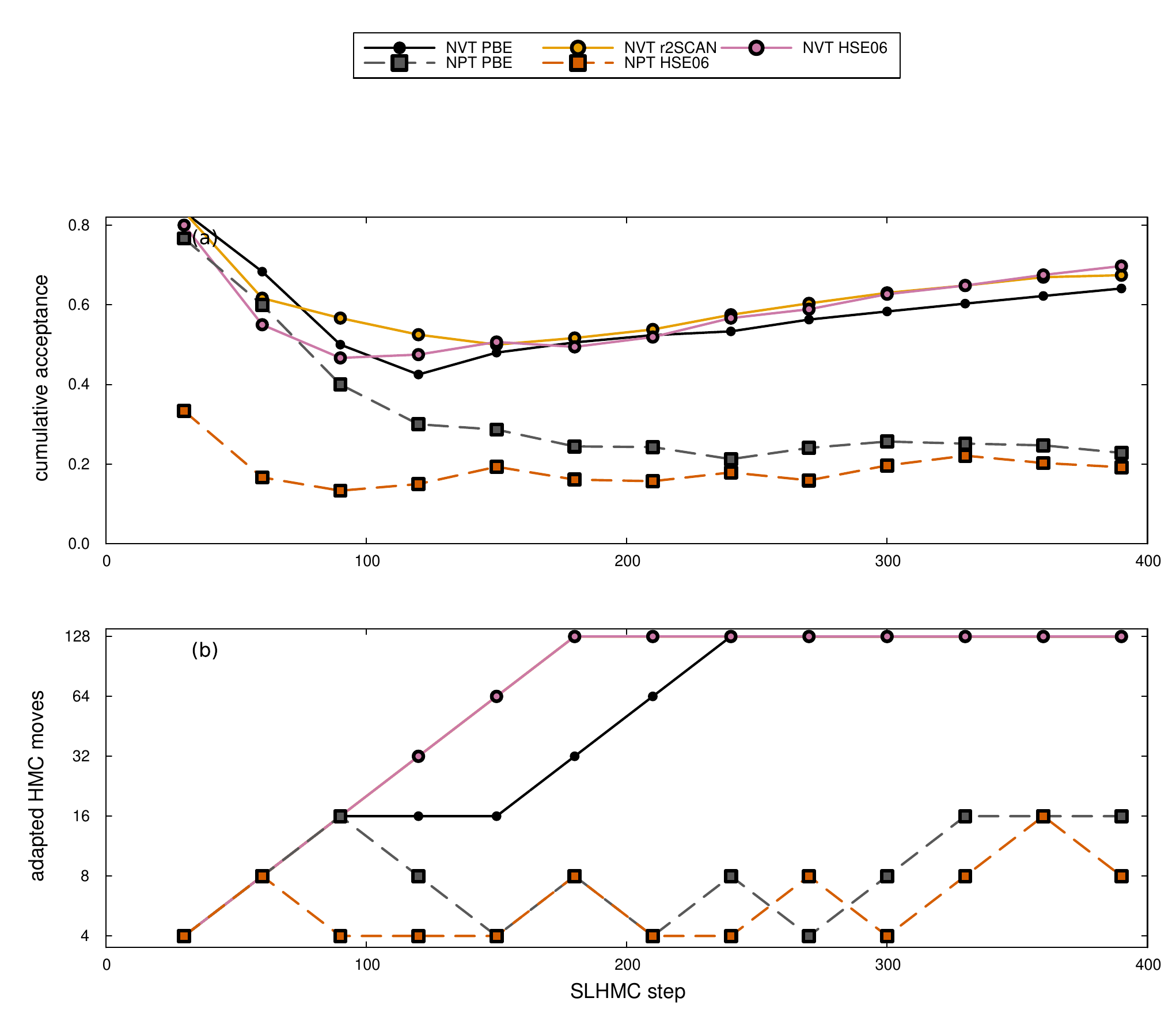}
\caption{Adaptive HMC proposal behavior during the SLHMC runs used elsewhere in the manuscript. (a) Cumulative acceptance ratio as a function of SLHMC step. (b) Adapted number of 1 fs MACE integration steps used for one HMC proposal. Solid lines denote NVT runs and dashed lines denote NPT runs.}
\label{fig:hmc-adaptation}
\end{figure}

Figure~\ref{fig:hmc-adaptation} summarizes how the HMC proposal length was adapted during the SLHMC runs that are directly used elsewhere in the manuscript. We therefore show only the NVT runs for PBE, r$^2$SCAN, and HSE06, together with the NPT runs for PBE and HSE06, to keep the comparison aligned with the main-text analyses. The cumulative acceptance ratios show that the NVT runs stabilize at relatively high acceptance, while the NPT runs settle to lower but still steady plateaus, reflecting the more demanding cell-fluctuation proposals. The lower panel shows that these acceptance levels were obtained through adaptive changes in proposal length rather than by fixing one HMC trajectory length throughout training, which is important because the proposal model itself changes after each fine-tuning update.

\end{appendices}

\clearpage


\bibliography{yn-bibliography}

\end{document}